\documentstyle[prb,aps,preprint]{revtex}
\newcommand{\afl}[2]{\frac{\partial #1}{\partial #2}}
\newcommand{\pE}{ev\afl{f^0_k}{\epsilon}}
\newcommand{\col}{\left(\frac{\partial f}{\partial t}\right)_{coll}}

\begin{document}
\draft
\title{Classical theory of the Hall-effect\\ in an inhomogeneous magnetic
field}
\author{Kasper Juel Eriksen and Per Hedeg\aa{}rd}
\address{\O{}rsted Laboratory, Niels Bohr Institute, \\
 Universitetsparken 5, DK-2100 Copenhagen, Denmark}
\date{\today}
\maketitle

\begin{abstract}
Inspired by recent experiments by Geim et al. we discuss the classical theory
of the Hall effect of a 2 dimensional electron gas
in an inhomogeneous magnetic field. The field modulation
is in the form of flux tubes created by a superconductor overlayer.
We find that an approach, where the vortices are treated as individual
scatterers contributing to the collision term in the Boltzman equation
will not work --- it leads to a vanishing Hall constant at $T=0$.
If the field is treated as a smooth contribution to the driving term in
the Boltzman equation, the classical Hall constant emerges, in agreement
with experiments when the Fermi wavelength is short in comparison with
all other lengths in the problem.
\end{abstract}

\section{Introduction}
We will in this paper discuss the Hall effect in the case, where the
applied magnetic field is spatially varying.  The theory will be based
on the Boltzman equation, and we shall only consider
the 2D case. Our
motivation comes from recent experiments by Andrei Geim et
al.\cite{geim}, who put a superconductor over a 2 dimensional electron
gas, and then measured the Hall voltage. The superconductor will only
allow the magnetic field to penetrate in Abrikosov vortices, thereby
modulating the field. For high fields the vortices are strongly
overlapping and the field is only slowly varying, so that the usual
Hall coefficient is to be expected. This is indeed what is seen
experimentally and it is the result of our calculations. For low fields
(below 100 G), where the vortices start to become spatially separated,
the Hall coefficient depends on the 2D electron density and therefore on
the de Broglie wavelength of the electrons at the Fermi surface.  When
the de Broglie wavelength is comparable to or greater than the
diameter of a vortex, the Hall effect is reduced by an almost field
independent fraction in low fields. The fraction is about $80 \%$
in the electron gas with the smallest density experimentally obtainable.
Since the effect depends on the electron
density one might expect that it is a quantum mechanical effect.  The
only way to put quantum mechanics into the Boltzmann equation is
through the scattering cross sections.
Khaetskii \cite{khaetskii} has
proposed treating the spatially separated vortices as asymmetric
scatterers. At de Broglie wavelengths much shorter
than the vortex diameter the electron will be scattered asymmetrically
and in accordance with a classical picture. At the opposite limit, first
treated by Aharonov and Bohm\cite{aharonov}, where the wavelength is much
larger than a vortex diameter the scattering is symmetric. Khaetskii and
earlier Kuptsov and Moiseev\cite{kuptsov} showed that the degree of
asymmetry gradually disappears as the diameter of the vortex is reduced
in comparison with the electron wavelength. Khaetskii's idea is that
this reduced asymmetric scattering can account for the reduced Hall
effect.
His calculations show that this is indeed possible in a
classical gas. We have extended his calculations to a degenerate gas
obeying Fermi-Dirac statistics. Here the Pauli principle, and its
restrictions on the scattering, will reduce the calculated Hall
coefficient to about $k_{\text{B}}T/\epsilon_{\text{F}}$, in strong
disagreement with the experiment.

We have subsequently considered the case where the
magnetic field is a slowly modulated field. This amounts to treating
the $B$-field as a driving force on the left hand side of the Boltzmann
equation.
Here we find the classical Hall effect corresponding to a homogeneous field
at all field strenghts and electron densities. This is to be
expected since we recover the experimental results when the electron
wavelength is shorter
than a vortex and we henceforth can talk about the magnetic field at the
electron's position with some confidence, while the procedure fails when the
electron wavelength is longer than the field modulations.

\section{Vortices as scattering centers}
First we will describe the magnetic flux tubes as independent
scattering centers, much like usual impurities. This of course is
supposed to apply only at the low field limit, where the tubes are
sufficiently far apart. The characteristic feature of scattering off
flux tubes is that the scattering probability is asymmetric: There
is an enhanced probability of electrons being scattered to the left.
We will model this by an asymmetric scattering probability
$w(k,\psi)$, where $\psi$ is the scattering angle and $k$ is the
length of the momentum vector.
The Boltzman equation, linearized in the external electric field, has
the familiar form
\begin{equation}
  -e\vec{v} \cdot\vec{E} \afl{f^0}{\epsilon} = \left (\frac{\partial
    f}{\partial t} \right)_{\text{coll}},
\label{boltzman1}
\end{equation}
where $f^0$ is the equibrilium distribution function. The collision
term consists of two parts, a flux tube part and a usual impurity
scattering part, which we will treat in the relaxation time
approximation.  Denoting the distribution function as $f(k,\phi)$,
$\phi$ being the angle between $\vec{k}$ and the external field $\vec{E}$, we
have
\begin{eqnarray}\label{stoed}
  \left (\frac{\partial f}{\partial t} \right)_{\text{coll}}&=& -
  \rho \int_{0}^{2\pi}\frac{d\psi}{2\pi}w(k,\psi)
  \bigl(f(k,\phi)(1-f(k,\phi+\psi)) -
  f(k,\phi-\psi)(1-f(k,\phi))\bigr)\nonumber \\ && \mbox{} -
  \frac{f(k,\phi)-f^0}{\tau}.
\end{eqnarray}
Here $\rho$ is the density of fluxtubes, i.e. $\rho=(B
A)/(\phi_0/2)/A=eB/h$.  The important difference to the work by
Khaetskii\cite{khaetskii}, is the inclusion of the Pauli principle.

We will solve the equation by Fourier transforming in the angle $\phi$.
Introducing
\begin{equation}
  f_n(k) = \int_{0}^{2\pi}\frac{d\phi}{2\pi}\, e^{in\phi} f(k,\phi),
  \qquad w_{n}(k) = \int_{0}^{2\pi}\frac{d\phi}{2\pi}\, e^{in\phi}
  w(k,\phi),
\end{equation}
the Boltzman equation (\ref{boltzman1}) has the form
\begin{eqnarray}  \label{fourierboltzman}
  \lefteqn{-evE \afl{f^0}{\epsilon} (\delta_{n,1}+\delta_{n,-1})/2 =
    -(f_n(k)-\delta_{n,0}f^0)/\tau} \\ && \mbox{}-\rho \Bigl(
  (w_{0}(k)-w_{n}(k))f_n(k)+\sum_m(w_{m}(k)-w_{-m}(k))f_{n-m}(k)f_m(k)\Bigr).
  \nonumber
\end{eqnarray}
The functions $w_{n}(k)$ satisfy $w_{n}(k)^*=w_{-n}(k)$. Accordingly there are
both real and imaginary contributions to the effective relaxation
time, due to the flux tubes. The imaginary parts have, as pointed out by
Khaetskii, a simple interpretation, namely as an effective homogeneous
magnetic field. Indeed in the Fourier transformed Boltzman equation with a
homogeneous magnetic field, the magnetic field term has the form:
\begin{equation}
  e(\vec{v}\times\vec{B})\afl{f(\vec{k})}{\vec{p}} \longrightarrow
  -in\omega_c f_n(k).
\label{magfield}
\end{equation}
Upon linearization of the last term in (\ref{fourierboltzman}) we get
$f_0=f^0$ and of the other terms only $f_1$ and $f_{-1}$ are non-vanishing and
they become
\begin{equation}
  f_1(k)=f_{-1}(k)^* = \frac{1}{2}\frac{e v \afl{f_0}{\epsilon}
    \tau(\epsilon)}{1 + i\rho(2f_0(k)-1)\text{Im}(w_1(k))\tau(\epsilon)} E,
  \label{solution}
\end{equation}
where
\begin{equation}
  \tau(\epsilon)^{-1} = \rho(1-\text{Re}(w_1(k)))+ \tau^{-1}.
  \label{taueffektiv}
\end{equation}
It is now straightforward to work out the conductivities. We get
\begin{equation}
  \sigma_{xx} = \frac{ne^2}{m} \left\langle
  \frac{\tau(\epsilon) }{1+((2f_0-1)\alpha(\epsilon))^2}
 \epsilon\left(-\afl{f_0}{\epsilon}\right) \right\rangle,
\label{conductivity}
\end{equation}
and
\begin{equation}
  \sigma_{xy} = -\frac{ne^2}{m} \left\langle
  \frac{\tau(\epsilon)\alpha(\epsilon) }{1+
    ((2f_0-1)\alpha(\epsilon))^2}
(2f_0-1)\epsilon\left(-\afl{f_0}{\epsilon}\right)\right\rangle,
  \label{hallconductivity}
\end{equation}
where the bracket $\langle\cdot\rangle$ is defined by
\begin{equation}
\langle A\rangle = \frac{\int d\epsilon A(\epsilon)}{\int d\epsilon
  f_0(\epsilon)},
  \label{bracket}
\end{equation}
and $\alpha(\epsilon)=\rho\tau(\epsilon)\text{Im}(w_1(\epsilon))$. Khaetskii
\cite{khaetskii} has shown that in the classical limit
$\alpha(\epsilon_{\text{F}}) = \omega_c \tau$.
The important difference between the result in equation
(\ref{hallconductivity})
and Khaetskii's result is the factor $2f_0(\epsilon)-1$, which is zero at the
Fermi level. This means that the Hall voltage will disappear at $T=0$.
The factor comes from the proper implementation of the Pauli
principle. In the low-$T$ ($T\ll\epsilon_{\text{F}}$)
limit, where we approximate $\alpha$ and $\tau$ by their value at the Fermi
level,
we simply get
\begin{equation}
\sigma_{xx} =
\frac{ne^2\tau(\epsilon_{\text{F}})}{m}\frac{\text{Arctan}
(\alpha(\epsilon_{\text{F}}))}
{\alpha(\epsilon_{\text{F}})}.
\end{equation}
By neglecting the term $((2f_0(\epsilon) -1)\alpha(\epsilon))^2$ in the
denominator in
(\ref{conductivity}) we get
\begin{equation}
\sigma_{xy} < \frac{k_{\text{B}}T}{\epsilon_{\text{F}}}
\frac{ne^2\tau(\epsilon_{\text{F}})}{m} \alpha(\epsilon_{\text{F}}).
  \label{lowT}
\end{equation}
In the experiment by Andrei Geim et al.\cite{geim} the temperature was
1.3 K.
If we use an effective mass of $0.07m_e$, $k_{\text{B}}T
/ \epsilon_{\text{F}}$ is less than 0.1. The electron mobilities were
in the range of 40-100 $\frac{\text{m}^2}{\text{V s}}$ and the
magnetic field was swept from 0 G to 200 G. Consequently $\alpha$ is
in the range of 0-2. In Figure 1 we have
plotted the Hall resistance (normalized to the classical
value $B_{\text{eff}}/ne$) as a function of the magnetic field. From here it is
seen that the Hall effect is reduced by a factor of about $k_{\text{B}}T
/ \epsilon_{\text{F}}$. To illustrate the crossover to the
non-degenerate electron gas case treated by Khaetskii we have
plotted the same quantity as a function of temperature in Figure 2.

The conclusion is that asymmetric scattering does not give rise to a
Hall effect in a degenerate electron gas. This result is not in
agreement with experiments. To explain this result we take
as a simple model the scattering to the left through the same angle $\phi_0$
at each scattering event.
\begin{equation}
w(\phi) \propto \delta(\phi -\phi_0)
\label{model}
\end{equation}
The effect of the $-e\vec{E}$ field is to make more electrons go
in it's direction ($\theta = 0$). The effect of the scattering and therefore
of the magnetic field is to oppose this effect by
scattering electrons out of the $\theta = 0$ direction.
In particular (considering only magnetic scattering)
\begin{equation}
\left (\frac{\partial f}{\partial t} \right)_{\text{mag-coll}} < 0
\label{stoedineq}
\end{equation}
{}From (\ref{stoed}) we have with our model scattering (\ref{model}) that
\begin{equation}
\left (\frac{\partial f}{\partial t} \right)_{\text{coll-mag}}(k,0)
= f(k,-\phi_0)(1-f(k,0)) -f(k,0)(1 - f(k,\phi_0))
\end{equation}
We want to determine the angle dependence of $f(k,\theta)$.
If $f(k,\theta)$ is small ($k>k_{Fermi}$) it is the angle dependence
of the  $f(k,\theta)$
outside the parentheses that dominates and we therefore drop the parentheses
and get from (\ref{stoedineq})
\begin{eqnarray}
f(k,-\phi_0) - f(k,0) & < & 0\\
f(k,-\phi_0) & < & f(k,0).
\end{eqnarray}
Consequently the electrons have a tendency to move to the left.
This classical picture is due to the fact that
the parentheses we have neglected are exactly  the contribution from
the Pauli principle.
If, on the other hand, $f(k,\theta)$ is close to $1$ ($k<k_{Fermi}$) the
Pauli contribution dominates and we consequently
drop the prefactor to the parentheses.
\begin{eqnarray}
(1-f(k,0)) - (1-f(k,\phi_0)) & < & 0\\
f(k,\phi_0) & < & f(k,0)
\end{eqnarray}
The electrons now tend to move to the right. The reason is
that in order to scatter in a dense Fermi gas it is essential that
there are few electrons a scattering angle away.
The electrons above and below the Fermi surface thus move in opposite
directions (the $1 - 2 f_0(\epsilon)$ factor) and the net
asymmetry is very small (it arises solely from the difference in speed
below and above the Fermi surface) --- the Hall effect has disappeared.

\section{Magnetic field as a driving force}
{}From the above section we conclude that it is not correct to treat
the vortices as scattering centers, that discontinously changes a
electron's position in phasespace. In this section we are going to discuss
a complementary approach, where we treat the inhomogeneous field as
a driving force, that changes a electron's position continously in phase space.
The approach is totally classical and is supposed only to apply
in a dense electron gas, where the electron de Broglie wavelength
is short in comparison with the length over which the magnetic field
varies. We will assume that the magnetic field is random, correlated over
lengths comparable to the effective London length.

To the usual linear order in the electric field
and in the deviation $g(\vec{r},\vec{k})$ from equilibrium $f^0(\vec{r},k)$,
the
Boltzmann equation is
\begin{equation}
\vec{v} \cdot \afl{g}{\vec{r}}(\vec{r},\vec{k}) - e\vec{E}(\vec{r})
\cdot\vec{v}
\afl{f^0}{\epsilon}(k)-e (\vec{v} \times \vec{B}(\vec{r})) \cdot
\afl{f}{\vec{p}}(\vec{r},\vec{k})
= \col(\vec{r},\vec{k}) .
  \label{boltz2}
\end{equation}
The collision contribution we will treat as scattering against fixed
impurities. Accordingly in polar coordinates in the $\vec{k}$-space
\begin{equation}
  \col(\vec{r},\theta) = \rho \int_{0}^{2\pi} \frac{d\phi}{2\pi}
w(\phi)(f(\vec{r},\theta-\phi) - f(\vec{r},\theta)),
  \label{collision}
\end{equation}
where $\rho$ is the density of scatterers. We have suppressed the $k$
dependence.

We will write the magnetic field as
$B(\vec{r} ) = B^0 + \delta B(\vec{r} )$, where $B^0$ is the average magnetic
field. The Boltzmann equation can now be written
\begin{eqnarray}
   v\cos{\theta} \afl{g}{x}(\vec{r},\vec{k}) + v\sin{\theta}
\afl{g}{y}(\vec{r},\vec{k})
   + \omega_c \afl{g}{\theta}(\vec{r},\vec{k})
     &&
\nonumber\\ \mbox{} +
 \frac{e \delta B }{m}\afl{g}{\theta}(\vec{r},\vec{k}) -
 \pE \cos{\theta} E_x(\vec{r})  - \pE \sin{\theta}E_y(\vec{r}) & = &
 \col(\vec{r},\vec{k})
\nonumber\\
  \label{boltz3}
\end{eqnarray}
with $\omega_c = eB^0/m$.

In (\ref{boltz3}) we introduce the hermitian operator
\begin{equation}
D  =  i \afl{}{\theta} + i r_c \left( \cos{\theta} \afl{}{x} + \sin{\theta}
\afl{}{y}
\right),
\end{equation}
where $r_c  =  v/\omega_{c}$ is the classical cyclotron radius in the magnetic
field $B^0$. We now want to simulate the effect of the collisions by a
relaxation time approximation, so we add and subtract a relaxation time
contribution and arrive at
\begin{eqnarray}
\left( D + \frac{i}{\omega_{c} \tau} \right) g & = &  \chi \\
& \equiv &  \frac{i}{\omega_{c}}\pE
(E_x\cos{\theta}+E_y\sin{\theta}) - i \frac{\delta B}{B^0}
\afl{g}{\theta}
\nonumber\\&&  \mbox{} +
i \frac{g}{\omega_{c}\tau} + \frac{i}{\omega_c} \col.\label{xhi}
\end{eqnarray}
Here $\tau$ can be chosen arbitrarily, but later the usual value of
$\tau$ will emerge as a natural choice.
The eigenfunctions of $D$ with eigenvalue $n$ ($n$ integer) are
\begin{equation}
\psi_{\vec{q},n} = \frac{1}{\sqrt{2\pi A}} \exp{(i \vec{q} \cdot \vec{r} -i n
\theta
-i r_c q_x \sin{\theta} + i r_c q_y \cos{\theta})},
\end{equation}
where $A$ is the area of the electron gas.
Therefore $g$ can be written
\begin{equation}
 g = \int d\vec{r}'\int d\theta'
G(\vec{r},\vec{r}',\theta,\theta')\chi(\vec{r}',\theta')
\end{equation}
with the Green function
\begin{eqnarray}
G(\vec{r},\vec{r}',\theta,\theta') & = &
\sum_{\vec{q},n}\frac{\psi_{\vec{q},n}(\vec{r},\theta)
\psi_{\vec{q},n}^{*}(\vec{r}',\theta')}{n+\frac{i}{\omega_{c}\tau}}\\
& = & \frac{1}{2\pi A} \sum_{\vec{q},n}\frac{e^{-i n(\theta - \theta')}}{n +
\frac{i}{\omega_c \tau}}
\exp{\biggl(i\vec{q} \cdot \bigl(\vec{r} -
\vec{P}_\theta(\theta-\theta',\vec{r}')\bigr)\biggr)}
\nonumber \\
& = &  \frac{1}{2\pi}\sum_n \frac{e^{-i n
(\theta -\theta')}}{n + \frac{i}{\omega_c\tau}}
\delta ( \vec{r} - \vec{P}_\theta(\theta-\theta',\vec{r}')),\nonumber
\end{eqnarray}
where
\begin{equation}
\vec{P}_\theta(\phi) = \vec{r} +r_c \left(
\begin{array}{c}
\sin{\theta}\\
-\cos{\theta}
\end{array}
\right) + r_c \left(
\begin{array}{c}
-\sin{(\theta - \phi)} \\
+\cos{(\theta-\phi)}
\end{array}
\right)
\end{equation}
 is the classical cyclotron orbit in the
homogeneous field $B^0$ parametrised by the momentum coordinates (angles).
We assume $\omega_c\tau$ is positive and get with $\phi = \theta -\theta'$,
using Poisson's summation formula
\begin{equation}
\frac{1}{2\pi} \sum_n \frac{e^{-in\phi}}{n + \frac{i}{\omega_c\tau}}
= \frac{i\exp{(-\frac{|\phi|}{\omega_{c}\tau})}}
{\exp{(-\frac{2\pi}{\omega_{c}\tau})}-1}
\end{equation}
where $|\phi|$ is the value of $\phi$ in $[0,2\pi[$. We, finally, have
\begin{equation}
g(\vec{r},\theta) = \frac{i}{\exp{(-2\pi/\omega_{c}\tau)}-1}\int_0^{2\pi}d\phi
\exp{(-\phi/\omega_{c}\tau)}\chi(\vec{P}_{\theta}(\phi),\theta-\phi)
\label{finalg}
\end{equation}
The physical interpretation of this formula is that you assume that the
electrons move along their
classical trajectory in a homogeneous magnetic field $B^0$. The correction to
the local electron density is obtained by adding field corrections from the
neighbouring points according to the number of electrons arriving from
neighbouring points to your fieldpoint.
The prefactor arises because we only integrate around the classical
circular orbit once --- we could drop it and instead integrate to
infinity.

For our later choice of relaxation time the
mean free path is very long (at least 2$\mu m$ and normally more than
10$\mu m$) so  in fact we make a field average along the classical trajectory.
Now the cyclotron radius $r_{c}$ is of the order 2$\mu m$  and
therefore much bigger than the magnetic correlation length $\xi$,
which is of the order 0.1$\mu m$, so
the system is strongly selfaveraging.

We now average and get
\begin{eqnarray}
<g> & = &  \frac{i}{e^{-2\pi/\omega_c\tau}-1}\int_0^{2\pi}d\phi
e^{-\phi/\omega_c\tau}<\chi(\vec{P}_\theta(\phi),\theta - \phi)>
\nonumber\\
& = & \frac{i}{e^{-2\pi/\omega_c\tau}-1}\int_0^{2\pi}d\phi
e^{-\frac{\phi}{\omega_c\tau}}\left(
i\afl{f^0}{\epsilon}\frac{ev}{\omega_c}(<E_x>\cos{(\theta-\phi)}
+<E_y>\sin{(\theta - \phi)})
\right.
\nonumber\\&& \mbox{} -
 \left.
\frac{i}{B^0}<\delta B \afl{g}{\theta}>(\theta-\phi) +
i\frac{<g>}{\omega_c\tau}(\theta -\phi)  +
\frac{i}{\omega_c} <\col(\theta - \phi)>\right)
\nonumber \\
& = & g^0(\vec{r},\theta,\tau) +
\frac{i}{e^{-2\pi/\omega_c\tau}-1}\int_0^{2\pi}d\phi
e^{-\frac{\phi}{\omega_c\tau}}\left( -
\frac{i}{B^0}<\delta B
\afl{g}{\theta}>(\theta-\phi)\right.
\nonumber\\ \mbox{} &&+
 \left. i\frac{<g>}{\omega_c\tau}(\theta -\phi) +
\frac{i\rho}{\omega_c} \int_0^{2\pi}\frac{d\phi'}{2\pi}
w(\phi')(<g(\theta-\phi-\phi')>-<g(\theta-\phi)> )\right) \nonumber
\end{eqnarray}
with
\begin{eqnarray}
g^0(\vec{r},\theta,\tau) = \tau\pE
\cos{\theta}\left(\frac{E^0_x}{1+(\omega_c\tau)^2}
-\frac{\omega_c\tau E^0_y}{1+(\omega_c\tau)^2}\right)
\nonumber\\ \mbox{} +
\tau\pE \sin{\theta}\left(\frac{\omega_c\tau E^0_x}{1+(\omega_c\tau)^2}
+\frac{E^0_y}{1+(\omega_c\tau)^2}\right).
\end{eqnarray}
Notice that $g^0$ is the distribution function in a homogeneous
magnetic field in the relaxation time approximation (with relaxation
time $\tau$).
We now decompose the angle part of the $\vec{k}$-space in Fourier components
\begin{eqnarray}
g(\theta) & = & \sum_n g_n e^{-in\theta}\\
w(\phi) & = & \sum_n w_n e^{-in\phi}
\end{eqnarray}
Integrating out $\phi$ and $\phi'$ we get
\begin{eqnarray}
<g(\theta)> &=&   g^0(\vec{r},\theta,\tau) + \sum_n\frac{i
e^{-in\theta}}{\frac{1}{\omega_c\tau}-in}
\frac{n<\delta B g_n>}{B^0} \nonumber \\
\mbox{} && + \sum_n \frac{<g_n> e^{-in\theta}}{\frac{1}{\omega_c\tau}-in}\left(
\frac{\rho(w_n -w_0)}{\omega_c} + \frac{1}{\omega_c\tau} \right)
\end{eqnarray}
The $n = 0$ Fourier component is trivial ($<g_0> = <g_0>$) and for $n\neq 0$ we
have
\begin{eqnarray}
<g_n> & = & g^0_n(\tau) + \frac{1}{\frac{1}{\omega_c\tau}-in}\left[
\frac{in<\delta B g_n>}{B^0} + \frac{\rho(w_n -1) +
  \frac{1}{\tau}}{\omega_c}<g_n> \right]\nonumber \\
& = & g^0_n(\tau) + \frac{1}{\frac{1}{\omega_c\tau}-in}\left[
\frac{in<\delta B\delta g_n>}{B^0} + \frac{\rho(w_n -1) +
  \frac{1}{\tau}}{\omega_c}<g_n> \right],
\end{eqnarray}
where we, in the last step, have used $<\delta B> = 0$.

The current is determined by $g_{1}$:
\begin{eqnarray}
j_x+ij_y & =& \int_0^\infty
\frac{kdk}{2\pi^2}\int_0^{2\pi}d\theta
\,g(k,\theta)(v\cos{\theta} + iv\sin{\theta})\\
& = & \int_0^\infty v\frac{kdk}{\pi}g_{1}.
\label{stroem}
\end{eqnarray}
 If we choose
\begin{equation}
\frac{1}{\tau} = \frac{1}{\tau_{1}} \equiv \rho( 1 -w_{1}) = \rho \int_0^{2\pi}
d\theta
(1-\cos{\theta}) w(\theta)
  \label{tau}
\end{equation}
we get
\begin{eqnarray}
<g_{1}> & = & g^0_{1}(\tau_{1})
+\frac{i}{\frac{1}{\omega_c\tau_{1}}-i}
<\frac{\delta B(\vec{r})}{B^0}\delta g_{1}(\vec{r})>.
\label{cortodetg}
\end{eqnarray}

In the appendix we have calculated the leading term in
$<\frac{\delta B(\vec{r})}{B^0}\delta g_{1}(\vec{r})>$
to be
\begin{eqnarray}
\frac{-i<g_{1}>}{e^{(-2\pi/\omega_{c}\tau_{1})}-1} \int_0^{2\pi}d\phi &&
e^{(-\phi/\omega_{c}\tau_{1})}e^{i\phi}<\frac{\delta B(\vec{r})}{B^0}
\frac{\delta B(\vec{P}_{\theta}(\phi))}{B^0}> \label{approx1}\\
&&\sim  <g_{1}> \frac{\xi}{r_c}<\frac{\delta B(\vec{r})}{B^0} \frac{\delta
  B(\vec{r})}{B^0}>   \\
&& =  <g_{1}> 0.06 \sqrt{\frac{10^{15}m^{-2}}{n}}.
\label{vurapprox1}
\end{eqnarray}
We use here that for randomly distributed gaussian vortices each
carring half a flux quantum, $<\frac{\delta B(\vec{r})}{B^0} \frac{\delta
  B(\vec{r})}{B^0}> = \frac{\hbar}{2 e\xi^2}$, $\xi = 0.1 \mu m$ and that $r_c
= \frac{522 G }{B}
\sqrt{\frac{n}{10^{15}m^{-2}}} \mu m$. In the experiment by Geim et
al. $\sqrt{\frac{10^{15}m^{-2}}{n}}$ varies between 0.5 and 1.7.
Therefore, the average electron distribution is  nearly as in the homogeneous
case. If we use the approximation (\ref{approx1}) in (\ref{cortodetg})
and assume that $\tau_1$ can be treated as a constant in the energy
integration in (\ref{stroem}) we get that
\begin{eqnarray}
\frac{\rho_{xy}}{B/ne} & = & 1 + \frac{1}{1 -
  e^{(-2\pi/\omega_{c}\tau_{1})}} \int_0^{2\pi}d\phi
e^{(-\phi/\omega_{c}\tau_{1})}\sin{\phi}<\frac{\delta B(\vec{r})}{B^0}
\frac{\delta B(\vec{P}_{\theta}(\phi))}{B^0}>
\label{hall}\\
\frac{\rho_{xx}}{m/ ne^2 \tau_1} & = & 1  \frac{\omega_c \tau}{1 -
  e^{(-2\pi/\omega_{c}\tau_{1})}} \int_0^{2\pi}d\phi
e^{(-\phi/\omega_{c}\tau_{1})}\cos{\phi}<\frac{\delta B(\vec{r})}{B^0}
\frac{\delta B(\vec{P}_{\theta}(\phi))}{B^0}>.
  \label{magneto}
\end{eqnarray}
The integral in (\ref{magneto}) is of the same order of magnitude as
the prefactor in (\ref{vurapprox1}), implying the magnetoresistance is
a few procent greater than it would have been in a homogeneous magnetic
field. Because the $\phi$-integration in (\ref{hall}) is restricted by the
correlation function to an interval from 0 to $\frac{\xi}{r_c}$, the
integal is about $\left(\frac{r_c}{\xi}\right)^2 \approx 400$ times smaller
than the
prefactor in (\ref{vurapprox1}). The approximation (\ref{approx1}) is therefore
not the dominant contribution to the deviations  in the Hall effect
from the homogeneous case. Consequently, the deviation from the
homogeneous Hall effect is at most a few promille. This
is in perfect agreement with the fact that the experiment by Geim et
al. showed no
deviations from the homogeneous result in a dense electron gas.

\section{Conclusion}
In the experiment by Geim et al.\cite{geim} a reduced
Hall effect in the Abrikosov vortex modulated field is only observed at
electron densities below $4\cdot 10^{15} m^{-2}$
and in a magnectic field of less than 100 Gauss. That is when the
external magnetic field varies appreciably within a de Broglie
wavelength of the electrons at the Fermi surface. In this regime it is
expected that the Boltzmann equation description breaks down, but
outside this regime our treatment of the vortices simply as a
modulated magnetic field in the Boltzmann equation agrees with the
experiment. To explain the reduced Hall effect one  has to
incorporate some kind of quantum mechanics. We have shown
that it is not feasable to describe the vortices as scatterers and
hide all the quantum mechanics in the calculation of the scattering
cross sections.

A full quantum treatment should certainly include multiple coherent
scattering by the vortices, because single scattering is contained in
the present Boltzman calculation. It is also clear that the Hall constant
is reduced, since in the limit of very thin vortices (or what amounts to
the same, a very dilute electron gas) the Hall constant will vanish. In
this limit the time symmetry breaking will vanish, because one can without
any change in the physics reverse the direction of the field by placing
an infinitely thin Dirac vortex carrying one flux quantum $h/e$ at each
of the external vortices that carries half a flux quantum; the Dirac vortices
having a field in the opposite direction of the external field.

We acknowledge discussions with Mads Brandbyge, Erland Brun Hansen,
Ayoe Hoff, Dung-Hai Lee,
Poul Erik Lindelof, Mads Nielsen  and Rafael Taboryski.

\appendix
\section{}

In this appendix we are going to calculate the correlation function
\begin{equation}
<\frac{\delta B(\vec{r})}{B^0}\delta g_{1}(\vec{r})> =
\frac{1}{2\pi}\int_0^{2\pi}e^{i\theta}<\frac{\delta B(\vec{r})}{B^0}
g(\vec{r},\theta)> d\theta.
\label{cor1}
\end{equation}
To do this we assume the higher order correlation functions factorize
the second order correlation function out and we henceforth have the
gausssian result
\begin{equation}
<\delta B(\vec{r})\Phi(B)> = \int d\vec{y} <\delta B(\vec{r}) \delta
B(\vec{y})> <\frac{\delta
  \Phi}{\delta B(\vec{y})}>.
  \label{gaussianapproximation}
\end{equation}
Using this in (\ref{cor1}) we
get that
\begin{eqnarray}
<\frac{\delta B(\vec{r})}{B^0}\delta g_{1}(\vec{r})> & = & \frac{1}{B^0}
\int_0^{2\pi}\frac{d\theta}{2\pi}
e^{i\theta} \int d\vec{y} <\delta B(\vec{r})\delta B(\vec{y})> <\frac{\delta
  g(\vec{r},\theta)}{\delta B(\vec{y})}>.
  \label{cor2}
\end{eqnarray}
Now we have from (\ref{finalg}) that
\begin{eqnarray}
\frac{\delta g(\vec{r},\theta)}{\delta B(\vec{y})}
&=& \frac{i}{\exp{(-2\pi/\omega_{c}\tau)}-1}\int_0^{2\pi}d\phi
\exp{(-\phi/\omega_{c}\tau)}\times
\nonumber\\&&
 \left\{ i\frac{ev}{\omega_{c}}\afl{f^0_k}{\epsilon}
(\frac{\delta E_x(\vec{P}_{\theta}(\phi))}{\delta B(\vec{y})}
\cos{(\theta-\phi)}+
\frac{\delta E_y(\vec{P}_{\theta}(\phi))}{\delta B(\vec{y})}
\sin{(\theta-\phi)})
\right.
\nonumber\\&& \mbox{}
- i \frac{\delta(\vec{y} -\vec{P}_{\theta}(\phi))}{B^0}
\afl{g}{\theta}(\vec{P}_{\theta}(\phi),\theta-\phi) - i \frac{\delta
B(\vec{P}_{\theta}(\phi))}{B^0}
\afl{}{\theta}\left(\frac{\delta
  g(\vec{P}_{\theta}(\phi),\theta-\phi)}{\delta B(\vec{y})}\right)
\nonumber\\&&  \mbox{} +
\left.  \frac{\delta}{\delta B(\vec{y})} \left( i \frac{g}{\omega_{c}\tau} +
\frac{i}{\omega_c} \col
\right)\right\} .
 \label{functionalg}
\end{eqnarray}
When the last term
is inserted in (\ref{cor2}) it is seen as before that if we choose
$\tau = \tau_{1}$ the term cancels. When we use this
expression below we will assume that this kind of cancellation can be
done, and erase this term. (\ref{functionalg}) is an equation to iteratively
determine
$\frac{\delta g(\vec{r},\theta)}{\delta B(\vec{y})}$ with the third term as
the driving term. We henceforth expand $<\frac{\delta B(\vec{r})}{B^0}
\delta g_{1}(\vec{r})>$ in
this term. The first order contribution is
\begin{eqnarray}
<\frac{\delta B(\vec{r})}{B^0}\delta g_{1}(\vec{r})> & = &
\frac{1}{\exp{(-2\pi/\omega_{c}\tau_{1})}-1}\int_0^{2\pi}d\phi
e^{(-\phi/\omega_{c}\tau_{1})}
\int_0^{2\pi}\frac{d\theta}{2\pi}
e^{i\theta} \int d\vec{y} \times
\nonumber\\&&
 <\frac{\delta B(\vec{r})}{B^0}\frac{\delta B(\vec{y})}{B^0}>  \delta(\vec{y}
 -\vec{P}_{\theta}(\phi))
\afl{<g>}{\theta}(\theta-\phi)
\nonumber\\
& = & \frac{1}{\exp{(-2\pi/\omega_{c}\tau_{1})}-1}\int_0^{2\pi}d\phi
\int_0^{2\pi}\frac{d\theta}{2\pi}
e^{i\theta} e^{(-\phi/\omega_{c}\tau_{1})} \times
\nonumber\\&&
 <\frac{\delta B(\vec{r})}{B^0}\frac{\delta B(\vec{P}_{\theta}(\phi))}{B^0}>
\afl{<g>}{\theta}(\theta-\phi).
  \label{zeroorder}
\end{eqnarray}
Now $<\frac{\delta B(\vec{r})}{B^0} \frac{\delta
B(\vec{P}_{\theta}(\phi))}{B^0}>$ only depends on the distance
between $\vec{r}$ and $\vec{P}_{\theta}(\phi)$. Consequently
\hbox{$<\frac{\delta B(\vec{r})}{B^0}\frac{\delta
B(\vec{P}_{\theta}(\phi))}{B^0}>$}  is independent of $\theta$ and we
can move the $\theta$-integral
through with the result that
\begin{eqnarray}
<\frac{\delta B(\vec{r})}{B^0} \delta g_{1}(\vec{r})> & =
&\frac{-i<g_{1}>}{\exp{(-2\pi/\omega_{c}\tau_{1})}-1} \int_0^{2\pi}d\phi
e^{(-\phi/\omega_{c}\tau_{1})}e^{i\phi}<\frac{\delta
B(\vec{r})}{B^0}\frac{\delta B(\vec{P}_{\theta}(\phi))}{B^0}>.
  \label{zeroresult}
\end{eqnarray}
$<\frac{\delta B(\vec{r})}{B^0}\frac{\delta B(\vec{P}_{\theta}(\phi))}{B^0}>$
is only large within a
correlation length $\xi$ and  it's size is estimated as
$<\frac{\delta B(\vec{r})}{B^0}\frac{\delta B(\vec{r})}{B^0}>$. Accordingly, as
an order of magnitude estimate we have
\begin{equation}
<\frac{\delta B(\vec{r})}{B^0}\delta g_{1}(\vec{r})>  \sim <g_{1}>
\frac{\xi}{r_c}<\frac{\delta B(\vec{r})}{B^0} \frac{\delta B(\vec{r})}{B^0}>.
  \label{magnitude}
\end{equation}
This is in our case much less than $<g_{1}>$.
To get the second order contribution we have to iterate
(\ref{functionalg}) once more, putting the driving term back into the first
two terms and the fourth term on the right hand side of
(\ref{functionalg}). We will first take the fourth term and here we
get
\begin{eqnarray}
<\frac{\delta B(\vec{r})}{B^0} \delta g_{1}(\vec{r})>_{\text{4'th term}}  = &&
\frac{1}{\exp{(-2\pi/\omega_{c}\tau_{1})}-1}
 \int_0^{2\pi}d\phi
e^{(-\phi/\omega_{c}\tau_{1})}
\int_0^{2\pi}\frac{d\theta}{2\pi}
e^{i\theta} \int d\vec{y} \times
\nonumber\\
<\frac{\delta B(\vec{r})}{B^0} \frac{\delta
  B(\vec{y})}{B^0} > <&&\frac{\delta B(\vec{P}_{\theta}(\phi))}{B^0}
\afl{}{\theta}\left(
\frac{1}{\exp{(-2\pi/\omega_{c}\tau_{1})}-1}\int_0^{2\pi}d\phi'
e^{(-\phi'/\omega_{c}\tau_{1})} \times \right.
\nonumber\\
\delta(\vec{y} -\vec{P}_{\theta}(\phi+\phi')&&) \left.
\afl{g}{\theta}(\vec{P}_{\theta}(\phi+\phi'),\theta-\phi-\phi')\right)>
\end{eqnarray}
\begin{eqnarray}
 &= &
\frac{1}{(\exp{(-2\pi/\omega_{c}\tau_{1})}-1)^2}\int_0^{2\pi}d\phi\int_0^{2\pi}d\phi'
e^{(-(\phi+\phi')/\omega_{c}\tau_{1})}
<\frac{\delta B(\vec{r})}{B^0} \frac{\delta
B(\vec{P}_{\theta}(\phi+\phi'))}{B^0} > \times
\nonumber\\&&
\int_0^{2\pi}\frac{d\theta}{2\pi}
e^{i\theta}
<\frac{\delta B(\vec{P}_{\theta}(\phi))}{B^0}\afl{^2
  g}{\theta^2}(\vec{P}_{\theta}(\phi+\phi'),\theta-\phi-\phi')>.
  \label{fourth}
\end{eqnarray}
The last integral is, apart from the differentiations, the same as the
original integral, just now spatially separated. Therefore it is not
greater than
$ <g_{1}> \frac{\xi}{r_c}<\frac{\delta B(\vec{r})}{B^0} \frac{\delta
B(\vec{r})}{B^0} >$ - the order of magnitude from before. Again the
correlation
function $<\frac{\delta B(\vec{r})}{B^0} \frac{\delta
B(\vec{P}_{\theta}(\phi+\phi'))}{B^0}>$ is only appreciable
within a distance of $\xi$. Since both $\phi$ and $\phi'$ are positive,
the two remaining integrals are restricted to a region of size
$\frac{\xi}{r_c}$. Implying that we have the following order of magnitude
estimate:
\begin{equation}
<\frac{\delta B(\vec{r})}{B^0} \delta g_{1}(\vec{r})>_{\text{4'th term}} \sim
\frac{\xi}{r_c}
\left(\frac{\xi}{r_c} <\frac{\delta B(\vec{r})}{B^0} \frac{\delta
B(\vec{r})}{B^0} >\right)^2 <g_{1}>.
  \label{andenstr}
\end{equation}
If  we use that the parenthesis is about 0.1 we get that this term
is  $10\frac{r_c}{\xi} \sim 10\frac{2 \mu m}{0.1 \mu m}$, i.e. 200 times
smaller than the first order contribution. To take care of the first two terms
in
(\ref{functionalg}) we use that
\begin{equation}
\frac{\delta E_x(\vec{P}_{\theta}(\phi))}{\delta B(\vec{y})} = \int d\vec{z}
\frac{\delta E_x(\vec{P}_{\theta}(\phi))}{\delta g_0(\vec{z})}  \frac{\delta
g_0(\vec{z})}{\delta B(\vec{y})}.
  \label{funcelectric}
\end{equation}
The last term is treated as above. We find that the first iterate
is 0. As explained in the main text the first order contribution to
$<\frac{\delta B(\vec{r})}{B^0}\delta g_{1}(\vec{r})>$ mainly influences the
magnetoresistance. Consequently, higher order terms
contribute significiantly to the Hall effect. The most important is  the first
term
that arises when you go beyond the gaussian approximation\footnote{In
  a magnetic field consisting of fluxtubes placed at random
\newline
\hbox{$
<\delta B(\vec{r}) \delta g_{1}(\vec{r})> = \sum_{n=1}^{\infty}
  \frac{N}{n!} \int d\vec{y}_1 \cdots d\vec{y}_n <\delta B(\vec{r}) \delta
B(\vec{y}_1)
  \cdots \delta B(\vec{y}_n)>_{1-flux}<\frac{\partial g_{1}(\vec{r})}{\partial
    B(\vec{y}_1) \cdots \partial B(\vec{y}_n)}>,$}
\newline
where $N$ is the number of fluxes. In this appendix we have calculated the
first term in
the sum. The higher order terms may be calculated in exactly the same manner. }
:
\begin{eqnarray}
&&
\frac{-<g_{1}>}{(\exp{(\frac{-2\pi}{\omega_{c}\tau_{1}})}-1)^2}\int_0^{2\pi}d\phi\int_0^{2\pi}d\phi'
e^{(-(\phi+\phi')/\omega_{c}\tau_{1})} e^{i(\phi+\phi')}
<\frac{\delta B(\vec{r})}{B^0} \frac{\delta B(\vec{P}_{\theta}(\phi))}{B^0}
\frac{\delta B(\vec{P}_{\theta}(\phi+\phi'))}{B^0} > . \nonumber\\
  &&\label{beygaus}
\end{eqnarray}
The order of magnitude of the relative deviation in the Hall effect
from the homogeneous case, due to (\ref{beygaus}), is $\left(0.06
\sqrt{\frac{10^{15}m^{-2}}{n}}\right)^2$. Implying that in a dense electron
gas the deviation in the Hall effect from the homogeneous case is about one
promille.

\begin{figure}
\caption{The theoretical Hall resistivity coming from treating the magnetic
fluxtubes as scatterers in the Boltzmann equation, normalized to the
classical homogeneous result $\frac{B}{ne}$ as a function of $\alpha =
\mu B$ at a temperature of 0.1, 0.5, 1.0 and 5.0 $\epsilon_{{\rm F}}$.
In the experiments by Andrei
Geim et al., who used mobilities $\mu$ in the range  of 40-100
$\frac{\text{m}^2}{\text{Vs}}$, they found in a dense gas the homogeneous
result
$\frac{B}{ne}$, at a temperature  of
less than 0.1 $\epsilon_{{\rm F}}$.}

\label{fig1}
\end{figure}

\begin{figure}
\caption{The hall resistivity normalized to the homogeneous value
  $\frac{B}{ne}$ as a function of the temperature for $\alpha = \mu B = 0
.5$.}

\label{fig2}
\end{figure}


\begin{references}
\bibitem{geim} A. K. Geim, S. J. Bending and I. V. Grigorieva, Phys. Rev.
Lett. {\bf 69}, 2252 (1992).
\bibitem{khaetskii} A. V. Khaetskii, J. Phys. C. {\bf 3}, 5515 (1991).
\bibitem{aharonov} Y. Aharonov and D. Bohm, Phys. Rev. {\bf 115}, 485 (1959).
\bibitem{kuptsov} D. A. Kuptsov and M. Yu. Moiseev, J. Phys. I France {\bf 1},
1165 (1991).
\end{references}
\end{document}